\documentclass[lettersize,journal]{IEEEtran}
\usepackage{amsmath,amsfonts}

\usepackage{array}
\usepackage[caption=false,font=normalsize,labelfont=sf,textfont=sf]{subfig}
\usepackage{textcomp}
\usepackage{stfloats}
\usepackage{url}
\usepackage{booktabs}
\usepackage{verbatim}
\usepackage{graphicx}
\usepackage{booktabs}
\usepackage{multirow}
\usepackage{verbatim}
\usepackage{booktabs,subcaption,amsfonts,dcolumn}
\usepackage{cite}
\usepackage{threeparttable}
\usepackage{amssymb}
\usepackage{algpseudocode}
\usepackage{amssymb}
\usepackage{tikz}
\usepackage{makecell}
\usepackage{algorithm}
\usepackage{algorithmicx}
\usepackage{algpseudocode}

\usetikzlibrary{positioning}
\usetikzlibrary{fit}
\usetikzlibrary{calc}
\begin{document}

\title{Weakly Supervised Tabla Stroke Transcription via an Adaptive Dynamic Rhythm Language Model (ADRM)}
\author{Rahul Bapusaheb Kodag, Vipul Arora 

\thanks{}
}
\maketitle

\begin{abstract}
Tabla Stroke Transcription (TST) is central to the analysis of rhythmic structure in Hindustani music, yet it remains challenging due to complex and dynamic rhythmic organization and the scarcity of strongly annotated data. Existing approaches largely rely on fully supervised learning with onset-level annotations, which are costly and impractical at scale. This work addresses TST in a weakly supervised setting, using only symbolic stroke sequences without temporal alignment of onsets. We propose a framework that combines a Connectionist Temporal Classification (CTC)-based acoustic model with a sequence-level rhythmic language model for rescoring, similar to that used in automatic speech recognition. The acoustic model produces a decoding lattice, which is refined using an Adaptive Dynamic Rhythm Language Model (ADRM) that combines $t\bar{a}la$-conditioned symbolic rhythmic regularities with local stroke dynamics. Moreover, we release a new performance-recorded tabla dataset, named \emph{Tabla Improvisation Dataset}, along with a complementary synthetic dataset for sequence-level weakly supervised TST. Experiments demonstrate consistent and substantial reductions in stroke error rates with ADRM compared to those with acoustic-only decoding, confirming the benefit of incorporating symbolic rhythmic regularities during lattice rescoring for accurate transcription.

\end{abstract}


\section{Introduction}
Automatic Music Transcription (AMT) aims to convert music audio into symbolic
representations such as musical scores or MIDI files and is a fundamental problem
in Music Information Retrieval (MIR)~\cite{D4}. While substantial progress has
been achieved for pitched instruments such as piano, guitar, and strings,
transcription of percussive instruments remains relatively underexplored
~\cite{D4, R6}. This gap is more evident in Indian art music, particularly
Hindustani music~\cite{meta}.

\par Tabla is one of the most widely used percussion instruments in Hindustani
music, providing the rhythmic foundation for both vocal and instrumental
performances. Beyond temporal structure, tabla playing also contributes to
expressive variation in performance. Tabla Stroke Transcription (TST) is
therefore important for analyzing the cyclic rhythmic structures
($t\bar{a}la$s) central to this tradition and for supporting pedagogical and
archival applications.

\par From a transcription perspective, TST is related to Automatic Drum Transcription (ADT) in Western music. Many recent ADT and TST systems rely primarily on fully supervised deep learning models~\cite{D4, S9, S10}. Such approaches require precise onset-level annotations, which are costly for tabla music and demand expert musicians and careful alignment. In this context, we refer to such onset-aligned training as strong supervision. Weak supervision, in contrast, uses only ordered stroke sequences, thereby enabling scalable data collection.

In our previous work~\cite{meta}, we explored meta-learning with limited onset-annotated data, but the approach still requires onset-level supervision. A more practical alternative is weakly supervised learning, which is well-suited to Hindustani music. The cyclic metrical structure of $t\bar{a}la$ provides an underlying
rhythmic framework within which stroke sequences are organized, making
sequence-level annotation both musically meaningful and feasible. This metrical framework is defined by the organization of $m\bar{a}tr\bar{a}s$ and cycle boundaries, whereas the temporal duration of a cycle depends on the performed tempo and the surface stroke realization may vary across performers and successive cycles.

\par  Weakly supervised sequence learning is well established in Automatic Speech Recognition (ASR), where Connectionist Temporal Classification (CTC) enables training without explicit alignment~\cite{CTC,ASR_review_2025}. However, such approaches remain largely unexplored for TST. 
Recent studies on tabla have primarily explored onset-based segmentation and subsequent stroke classification~\cite{S12,1_gowriprasad2020onset}, while recent unsupervised methods learn acoustically meaningful stroke representations without requiring manual alignment~\cite{1_4_gowriprasad2025unsupervised}. In general, providing target information during training can improve task-specific learning compared with learning from the input data alone. For TST, the ordered stroke sequence provides useful target information while being substantially easier to annotate than precise onset-level annotations, as it does not require precise temporal localization of each stroke. We therefore formulate TST as a weakly supervised sequence transcription problem, in which the conventional symbolic tabla stroke sequence (e.g., $Dha, Dhin, Tin, Na$) is learned without explicit onset labels.

ASR systems also improve acoustically generated hypotheses through language-model-based rescoring~\cite{ASR_error_correction_TR_2025}, and related music language models have been used for music transcription for pitched instruments~\cite{MLM_2021}.  Rhythmic and repetitive structure has also been exploited for percussion-pattern discovery and musical segmentation~\cite{S8,sankaran2015}. In contrast, our goal is complete symbolic stroke transcription from audio using sequence-level supervision, with rhythmic information incorporated directly during CTC lattice decoding.
We therefore study weakly supervised TST using symbolic stroke sequences and use the proposed Adaptive Dynamic Rhythm Language Model (ADRM) to incorporate rhythmic information during CTC lattice decoding.

{The key contributions of this work are as follows:
\begin{enumerate}

\item We formulate Tabla Stroke Transcription as a weakly supervised sequence modelling problem based on symbolic stroke sequences and establish a benchmark setting for sequence-level TST in Hindustani music.

\item We curate and release a publicly available tabla dataset specifically designed for sequence-level weakly supervised TST, comprising 133 minutes of real-world theka-centered recordings with improvisatory variations across six commonly used $t\bar{a}la$s, together with a 121-minute synthetic dataset covering the same rhythmic structures.

\item We propose an Adaptive Dynamic Rhythm Language Model (ADRM) that combines global rhythmic patterns with local improvisatory variations for rhythm-aware lattice rescoring, leading to substantial improvements in transcription accuracy.
\end{enumerate}}
The remainder of the paper is organized as follows. Section~\ref{literature} reviews related work, Section~\ref{sec_proposed_system} presents the proposed method, Section~\ref{curated_dataset} describes the datasets, Section~\ref{sec_exp_setup} presents the experimental setup, and Section~\ref{sec_results} discusses the results. Section~\ref{sec_conclusion} concludes the paper.

\section{Related Work}\label{literature}
\subsection{Rhythmic Structure and Tabla Stroke Transcription}
In Hindustani music, a $t\bar{a}la$ defines the cyclic rhythmic framework of a composition and is realized through characteristic tabla stroke patterns. Each $t\bar{a}la$ has a canonical stroke sequence, known as the $thek\bar{a}$, which provides a rhythmic reference, while performances may include compositional and improvisatory variations. We focus on six commonly used $t\bar{a}la$s: $Dadr\bar{a}$, $Ekt\bar{a}l$, $Jhapt\bar{a}l$, $Keherv\bar{a}$, $R\bar{u}pak$, and $T\bar{i}nt\bar{a}l$. Detailed musicological
discussions of $t\bar{a}la$ are provided in~\cite{G4, G5}, while a stroke-level
illustration of the canonical $thek\bar{a}$ and its local improvisational
variations is given in Supplementary Material (S. VI).

TST aims to identify and label tabla strokes from audio. Early approaches used segment-and-classify pipelines, while later work adopted end-to-end deep learning models \cite{meta,S9,S10,S12}, similar to ADT\cite{D4}. Tabla studies have addressed onset detection\cite{1_gowriprasad2020onset}, structural segmentation \cite{4_gowriprasad2022structural} and gharana recognition\cite{4_gowriprasad2021tabla}, and unsupervised clustering based on acoustic and timbral similarity \cite{1_4_gowriprasad2025unsupervised}.

Related work in Carnatic music has also explored mridangam stroke transcription~\cite{mridang} and automatic composition segmentation using repeated musical structure \cite{sankaran2015}. Such repetition-based cycle information can provide useful rhythmic cues, but adapting this idea to Hindustani tabla requires a separate formulation because the cited work is developed primarily around Carnatic composition/Pallavi structure. Moreover, stroke-count-based $t\bar{a}la$ inference requires a reliable stroke sequence, whereas the objective of the present work is to obtain that stroke sequence from audio.

Gupta et al.'s \cite{S8} transcription serves as an intermediate step for retrieving selected recurring patterns in a known rhythmic setting, whereas our rhythmic model is integrated directly into CTC lattice rescoring rather than used for pattern retrieval. To the best of our knowledge, prior CTC-based weakly supervised drum transcription has primarily focused on Western drum transcription under constant-tempo assumptions~\cite{CTC_ADT}.

\subsection{Weakly Supervised Sequence Learning in ASR}

Weakly supervised sequence learning is widely used in Automatic Speech Recognition (ASR), where alignment-free objectives such as Connectionist Temporal Classification (CTC) enable training without frame-level annotations~\cite{CTC,ASR_review_2025}. Recognition accuracy is further improved through sequence-level refinement, including language model rescoring and error correction~\cite{ASR_error_correction_TR_2025}, demonstrating the value of constraints beyond acoustic evidence.

\subsection{Music Language Modelling for Transcription}
Inspired by language modelling in ASR, music language modelling (MLM) captures symbolic musical structure, such as harmony, rhythm, and temporal consistency, and integrates it with acoustic models for transcription. Existing MLM research has primarily focused on strongly supervised transcription of Western instruments, particularly piano, using probabilistic and neural sequence models~\cite{MLM_2021,MLM_2024}. Extensions to non-Western music remain limited~\cite{MLM_2025}.
Rhythmic pattern modelling has also been explored in tabla~\cite{S8};
However, weakly supervised rhythm-aware percussion transcription remains
less explored. The present work integrates rhythmic structure directly into
CTC lattice decoding for weakly supervised TST.

\section{Proposed Method}\label{sec_proposed_system}
This section presents the proposed framework for TST, which converts a tabla audio recording into a sequence of discrete stroke symbols. A CTC-based acoustic model first generates a decoding lattice containing multiple acoustically plausible stroke sequences, which are then refined using the proposed Adaptive Dynamic Rhythm Language Model (ADRM). ADRM combines a global rhythmic prior learned from symbolic training data with a local adaptive model that captures short-term, performance-specific stroke transitions. The following
subsections describe the acoustic model formulation, the components of
ADRM, and the lattice-based rescoring procedure.

\subsection{Acoustic Model}\label{acoustic_model}
Let $X \in \mathbb{R}^{F \times T}$ denote the acoustic feature matrix, with $F$
features extracted over $T$ time frames, and let
$Y = (y_1,\dots,y_U) \in \mathcal{S}^U$ denote the corresponding stroke sequence
of length $U \le T$, where $\mathcal{S}$ is a finite vocabulary of stroke types.
The acoustic model is trained using the Connectionist Temporal Classification
(CTC) loss~\cite{CTC},
\begin{equation}\label{eq:CTC}
L_{\mathrm{CTC}}(X,Y) = -\log P(Y \mid X),
\end{equation}
where the sequence likelihood $P(Y \mid X)$ marginalizes over all valid
frame-level alignments consistent with the target sequence $Y$.
\par After training, beam search generates a CTC decoding lattice containing multiple acoustically plausible stroke sequences. This lattice serves as the input to the rhythmic rescoring framework
described next.
\subsection{Adaptive Dynamic Rhythm Language Model (ADRM)}
ADRM operates on the CTC decoding lattice, assigning rhythm-aware scores to candidate stroke sequences. Let
$\mathcal{L} = (V, E, w_{\mathrm{ac}})$ denote the decoding lattice, where $V$ is
the set of nodes, $E$ the set of directed arcs, and $w_{\mathrm{ac}}(e)$ the
acoustic log-score associated with arc $e \in E$. Each path through $\mathcal{L}$ corresponds to a candidate stroke sequence
$\mathbf{s} = (s_1,\dots,s_K)$, with $s_k \in \mathcal{S}$. During decoding, ADRM updates the scores of candidate paths in the lattice using learned stroke-sequence regularities and short-term stroke transitions. It uses two complementary models: a global rhythmic prior that captures $t\bar{a}la$-conditioned symbolic stroke-sequence regularities learned from the training corpus, and a local adaptive rhythm model that captures short-term, path-dependent transition patterns.

The global rhythmic prior is learned from the available symbolic training
data and captures statistical rhythmic regularities rather than enforcing a
fixed stroke pattern for a particular $t\bar{a}la$. Thus, the global prior represents probabilistic stroke-sequence regularities associated with a $t\bar{a}la$ rather than a fixed temporal realization, fixed stroke count, or canonical $thek\bar{a}$ that must be followed in every cycle. Candidate sequences
receive rhythmic scores according to these learned regularities, while the
local adaptive component updates its transition statistics from the observed
stroke sequence to accommodate performance-specific and previously unseen
local variations.

\subsubsection{Global Rhythmic Prior Model}
\label{static_prior}
The global prior captures statistical dependencies among stroke symbols conditioned on the $t\bar{a}la$, without explicitly representing $m\bar{a}tra$ position, $s\bar{a}am$, $vibh\bar{a}g$, cycle phase, or cycle boundaries. Although tabla performance can exhibit substantial
compositional and improvisatory variation, the proposed framework captures
recurring sequence regularities observed in the training corpus. We capture these regularities using a global rhythmic prior, which assigns probabilities to stroke sequences based on $t\bar{a}la$-conditioned symbolic stroke statistics.
The proposed framework does not assume a single functional form for the
global rhythmic prior and supports multiple realizations within the same
decoding architecture.
In this work, we consider two realizations: a Bayesian $n$-gram model with
analytically estimated transition statistics and a Neural Next-Stroke
Language Model trained on symbolic stroke sequences.

\paragraph{Bayesian Global Rhythmic Prior}
The Bayesian formulation models $t\bar{a}la$-conditioned symbolic stroke-sequence regularities using an n-gram representation with analytically estimated stroke-transition statistics.
Because characteristic stroke-sequence statistics differ across tālas, we condition the global rhythmic prior on the underlying $t\bar{a}la$.

Let $\mathcal{T}$ denote the finite
set of $t\bar{a}la$s observed in the training data, and let $\tau \in \mathcal{T}$
be a specific $t\bar{a}la$. 
During training, $t\bar{a}la$ labels are used to estimate a separate
$t\bar{a}la$-conditioned global rhythmic prior for each $\tau \in \mathcal{T}$:
$P_{\mathrm{prior}}(s_k \mid \tau, s_{1:k-1})$. To avoid data sparsity from long stroke histories, we limit the effective
context to a recent stroke window $\mathbf{u}_k=s_{k-W:k-1}$ when estimating
the Bayesian global rhythmic prior.

For multi-$t\bar{a}la$ recordings, the underlying $t\bar{a}la$ is not
provided at test time. Rather than selecting a single $t\bar{a}la$ in
advance, we treat it as a latent variable and marginalize over the possible $t\bar{a}las$ to obtain a mixture of $t\bar{a}la$-conditioned global rhythmic priors.. Concretely, we introduce a
latent $t\bar{a}la$ variable $\tau^{(k)} \in \mathcal{T}$ at stroke position
$k$ and marginalize over it as:
\begin{equation}\label{eq:P_prior_TI}
\begin{aligned}
P_{\mathrm{prior}}^{(\mathrm{TI})}(s_k \mid \mathbf{u}_k)
&= \sum_{\tau \in \mathcal{T}}
   P(\tau^{(k)}=\tau \mid \mathbf{u}_k) \\
&\quad \times P_{\mathrm{prior}}(s_k \mid \tau, \mathbf{u}_k),
\end{aligned}
\end{equation}
For notational simplicity, we write
$P(\tau \mid \mathbf{u}_k)$ instead of
$P(\tau^{(k)}=\tau \mid \mathbf{u}_k)$.

\paragraph*{Estimation of the latent $t\bar{a}la$ Posterior}
The posterior $P(\tau \mid \mathbf{u}_k)$ is inferred during decoding and
controls the contribution of each $t\bar{a}la$-specific model to the global
rhythmic prior as stroke evidence accumulates. The posterior is then estimated
as
\begin{equation}\label{eq:P_tau_h}
P(\tau \mid \mathbf{u}_k)
= \frac{C(\tau, \mathbf{u}_k)\, P(\tau)}
       {\sum_{\tau' \in \mathcal{T}} C(\tau', \mathbf{u}_k)\, P(\tau')},
\end{equation}
where $C(\tau, \mathbf{u}_k)$ denotes the number of occurrences of window
$\mathbf{u}_k$ in training sequences labeled with $t\bar{a}la$ $\tau$, and
$P(\tau)$ is a prior proportional to the frequency of $t\bar{a}la$ $\tau$ in the
training data. Thus, for multi-$t\bar{a}la$ recordings, the decoder infers a posterior
distribution over candidate $t\bar{a}la$s from the observed stroke history,
without requiring the $t\bar{a}la$ label as an input.

\paragraph{Neural Global Rhythmic Prior}\label{sec:NN_LM}
As an alternative to the Bayesian formulation, we introduce a Neural
Next-Stroke Language Model (NN-LM) that predicts the next stroke from the preceding symbolic sequence. Unlike the analytical formulation, the NN-LM does not
explicitly condition its predictions on $t\bar{a}la$; instead, it learns symbolic stroke-sequence regularities implicitly from the training data.
Given a stroke prefix $s_{1:k-1}$, the
model parameterized by $\theta'$ outputs
\begin{equation}\label{eq_P_NN}
P_{\mathrm{NN}}(s_k \mid s_{1:k-1};\theta').
\end{equation}
Training minimizes the negative log-likelihood over a corpus of stroke
sequences,
\begin{equation}
\mathcal{L}_{\mathrm{NN}}(\theta')
= - \sum_{i=1}^{N} \sum_{k=1}^{K_i}
    \log P_{\mathrm{NN}}(y_{i,k} \mid y_{i,1:k-1}; \theta'),
\end{equation}
where $Y_i = (y_{i,1},\dots,y_{i,K_i})$ denotes the $i$-th ground-truth stroke
sequence. During decoding, the NN-LM directly replaces the Bayesian
global rhythmic prior
$P_{\mathrm{prior}}^{(\mathrm{TI})}(s_k \mid \mathbf{u}_k)$ in the ADRM,
while the local adaptive model remains unchanged.

\subsubsection{Local Adaptive Rhythm Model with Temporal Forgetting}\label{sec:dynamic_model}
A global canonical model is not sufficient because of local repetition and
improvisation. Hence, we incorporate a local adaptive rhythm model that models short-term rhythmic regularities and updates its transition statistics based on recently observed stroke sequences.

For this, we use a Dirichlet-parameterized first-order Markov model with exponentially weighted updates. This allows simple closed-form updates of transition probabilities and yields an interpretable representation.
For a stroke transition \( r \rightarrow q \), where \( r,q \in \mathcal{S} \),
we model the transition probability using a Dirichlet prior parameterized by
pseudo-counts \( \alpha = \{\alpha_{r,q}\} \), where \( \alpha_{r,q} \) denotes
the pseudo-count associated with the transition \( r \rightarrow q \).

\paragraph{Temporal Adaptation with Exponential Forgetting.}
Given a candidate lattice path, the model updates its transition statistics
based on recently observed stroke transitions. To remain responsive to local
rhythmic variations while preventing the unbounded growth of counts, we employ
exponential forgetting. After observing a transition
$s_{k-1}=r \rightarrow s_k=q$, the Dirichlet parameters are updated as
\begin{equation}
\label{eq:dirichlet_update}
\alpha_{r,q}^{(k)}
= (1-\rho)\,\alpha_{r,q}^{(k-1)}
+ \rho\,\delta(s_{k-1}=r, s_k=q),
\end{equation}
where $\rho \in (0,1)$ controls the effective memory timescale and
$\delta(\cdot)$ is 1 when the transition $r \rightarrow q$ occurs and 0
otherwise.  This update corresponds to an exponentially weighted moving average
over recent transitions. The resulting next-stroke distribution is defined as
\begin{equation}
\label{eq:P_adaptive}
{P_{\mathrm{adaptive}}}(s_k = q \mid s_{k-1}=r)
= \frac{\alpha_{r,q}^{(k)}}{\sum_{s \in \mathcal{S}} \alpha_{r,s}^{(k)}}.
\end{equation}
Although the parameters $\alpha^{(k)}$ evolve over time and implicitly encode 
past history, the predictive distribution ${P_{\mathrm{adaptive}}}$ remains first-order 
Markov, depending only on the immediately preceding stroke $s_{k-1}$.

\paragraph{Initialization of Dirichlet Parameters}
The adaptive model is initialized using global stroke-transition statistics
computed from the training data. Let $C(r \rightarrow q)$ denote the total number
of observed transitions from stroke $r$ to stroke $q$ across all training
sequences. The initial parameters are set as
\begin{equation}
\label{eq:alpha_init}
\alpha_{r,q}^{(0)} = C(r \rightarrow q) + \varepsilon_{\mathrm{dir}},
\end{equation}
where {$\varepsilon_{\mathrm{dir}} = 1.0$} is a small smoothing constant. For the special
start symbol $\langle s \rangle$ (representing sequence 
beginning), we use a uniform initialization
$\alpha_{\langle s \rangle,q}^{(0)} = 1$ for all $q \in \mathcal{S}$. 


\subsubsection{Combined Rhythm Language Model}
\label{sec:combined_rhythm}
The global rhythmic prior model and the local adaptive rhythm model are combined
linearly using an interpolation weight $\lambda \in [0,1]$, which controls
their relative influence. Mathematically, the probability of the next stroke during lattice rescoring is obtained as

\begin{equation}
\label{eq:P_comb}
\begin{aligned}
P_{\mathrm{comb}}(s_k \mid \mathbf{u}_k)
&= (1-\lambda)\,
   P_{\mathrm{prior}}^{(\mathrm{TI})}(s_k \mid \mathbf{u}_k) \\
&\quad +\;
   \lambda\,
   {P_{\mathrm{adaptive}}}(s_k \mid s_{k-1}).
\end{aligned}
\end{equation}
The sequence-level rhythmic score for a candidate stroke sequence
$\mathbf{s} = (s_1,\dots,s_K)$ is then computed as
\begin{equation}
\label{eq:log_P_comb}
\log P_{\mathrm{comb}}(\mathbf{s})
= \sum_{k=1}^{K}
  \log P_{\mathrm{comb}}(s_k \mid \mathbf{u}_k).
\end{equation}

\subsection{Lattice-Based Rhythmic Rescoring}
\label{sec:lattice_rescoring}
Rhythmic rescoring is performed offline on the CTC decoding lattice. The acoustic model first generates the complete lattice, which is then rescored using the proposed rhythmic model.
Decoding is carried out
offline: the acoustic model first generates a complete CTC lattice, which is then
rescored using rhythmic information.

In a standard CTC lattice, multiple paths with different stroke histories may
merge at the same lattice node. While such merging is appropriate for purely
acoustic scoring, it is problematic for rhythmic models, whose probabilities
depend explicitly on the preceding stroke sequence. Consequently, different
paths reaching the same node may require different rhythmic scores for subsequent
stroke extensions.

We address this issue using a state expansion strategy, which preserves
path-specific rhythmic context by treating identical lattice nodes reached via
different stroke histories as distinct decoding states.The output lattice $\mathcal{L}_{\mathrm{out}}$ therefore maintains multiple expanded states 
$(v,\alpha_1), (v,\alpha_2), \ldots$ at a single lattice node $v$, enabling correct rhythmic 
probability computation despite acoustic path merging.

\subsubsection{Expanded Decoding State}
\label{sec:expanded_state}
Each expanded decoding state is a tuple $(v,\alpha)$ where node $v$ is augmented 
with Dirichlet parameters $\alpha$ encoding rhythmic context through observed 
stroke transitions. Each state implicitly maintains the complete path history from the start, which enables unambiguous backtracking at the end of decoding.

\paragraph{State Representation and Initialization.}
To ensure consistent initialization of the adaptive rhythm model, we prepend a 
special start symbol $\langle s \rangle$ to each stroke sequence. The initial 
expanded state is $(v_0,\alpha^{(0)})$ with uniform initialization 
$\alpha^{(0)}_{\langle s \rangle,q}=1$ for all $q\in\mathcal{S}$, allowing the 
model to learn transitions from the first position without bias.
Let $k \in \{1, 2, \ldots, K\}$ denote the decoding position (stroke index).
At position $k$, the immediate predecessor stroke $s_{k-1}$ is the last observed
stroke, retrieved from the incoming arc label to node $v$.

 \subsubsection{Expanded Lattice Construction with Adaptive Rhythm Weighting}
Algorithm~\ref{algo_1} summarizes the proposed lattice rescoring procedure. Each outgoing arc is rescored by combining acoustic and rhythmic evidence. Rather than using a fixed rhythmic weight, we employ an \emph{adaptive rhythm weighting} scheme in which the contribution of rhythmic information depends on the reliability of the acoustic evidence.

At each decoding step $k$, the lattice node $v$ has multiple outgoing arcs for valid next strokes. The acoustic model assigns log-scores to these candidates. Acoustic confidence is quantified via an entropy-based measure that reflects how clearly one candidate dominates over others.

Let $\tilde{p}(e)$ denote the softmax-normalized acoustic scores 
for arc $e$ from node $v$:

\begin{equation}
\tilde{p}(e) = \frac{\exp(w_{\mathrm{ac}}(e))}
                     {\sum_{e' \in \text{Out}(v)} \exp(w_{\mathrm{ac}}(e'))}
\end{equation}
where $w_{\mathrm{ac}}(e)$ is the acoustic log-score. The entropy 
over these competing candidates is:

\begin{equation}
H(\tilde{p}) = -\sum_e \tilde{p}(e) \log \tilde{p}(e)
\end{equation}
Acoustic confidence at position $k$ is defined as:
\begin{equation}
\label{eq:confidence}
C_k = 1 - \frac{H(\tilde{p})}{\log |Out(v)|}
\end{equation}
High values of $C_k$ indicate confident predictions dominated by a single arc, while low values indicate ambiguity among competing stroke hypotheses.

The adaptive rhythm weight is defined as $\beta_k = \beta (1 - C_k)$. 
As a result, rhythmic information has greater influence when the 
acoustic evidence is ambiguous and less influence when the acoustic 
evidence is reliable.

For each expanded state $(v, \alpha)$ at position $k$, we rescore the outgoing arcs 
from node $v$. For an outgoing arc $e = (v \to v')$ labeled with stroke $q$, the 
algorithm computes the dynamic rhythmic
probability ${P_{\mathrm{adaptive}}}(\cdot \mid s_{k-1})$ (Eq.~\ref{eq:P_adaptive}) and the
global rhythmic prior $P_{\mathrm{prior}}^{(\mathrm{TI})}(\cdot \mid \mathbf{u}_k)$
(Eq.~\ref{eq:P_prior_TI}). These components are combined to
form the rhythmic probability $P_{\mathrm{comb}}(q \mid \mathbf{u}_k)$ using
Eq.~(\ref{eq:P_comb}). The arc is then rescored as
\begin{equation}
\label{eq:score_e}
\mathrm{score}(e)
= w_{\mathrm{ac}}(e) + \beta_k \log P_{\mathrm{comb}}(q \mid \mathbf{u}_k),
\end{equation}
where $w_{\mathrm{ac}}(e)$ denotes the acoustic log-score and $\beta_k$ controls the
contribution of rhythmic information at decoding position $k$.
After rescoring, the Dirichlet parameters $\alpha$ are updated according to
Eq.~(\ref{eq:dirichlet_update}), yielding the updated state $\alpha'$. The
resulting expanded state $(v',\alpha')$ is added to the output lattice
$\mathcal{L}_{\mathrm{out}}$ and inserted into the priority queue $Q$ for further
expansion.

\subsubsection{Search Strategy and Beam Pruning}
\label{sec:search_beam}
Expanded decoding states are explored using a best-first traversal, ordered by
their accumulated log-scores. States $(v,\alpha)$ are maintained in a priority
queue $Q$, initialized with the start state $(v_0,\alpha^{(0)})$.
To maintain computational tractability, beam pruning is applied at each expansion
step. Let $\mathrm{score}(v,\alpha)$ denote the accumulated score of a
state in $Q$. After expanding the current state and inserting successor states,
the maximum score in $Q$ is computed:
\begin{equation}
\label{eq:score_max}
\mathrm{score}_{\max}
= \max_{(v,\alpha)\in Q}
\mathrm{score}(v,\alpha).
\end{equation}
Only states within a beam width $\Delta_{\mathrm{beam}}$ of the best hypothesis are retained:
\begin{equation}
\label{eq:beam_p}
Q \leftarrow \left\{(v,\alpha)\in Q :
\mathrm{score}(v,\alpha)
\ge \mathrm{score}_{\max} - \Delta_{\mathrm{beam}}\right\}.
\end{equation}
This retains the most promising hypotheses while pruning unlikely paths.

\begin{algorithm}[h]
\caption{Adaptive Rhythmic Rescoring via State Expansion
in a CTC Decoding Lattice}
\label{algo_1}
\begin{algorithmic}[1]
\Require
CTC lattice $\mathcal{L}_{\mathrm{in}} = (V, E, w_{\mathrm{ac}})$ with 
  start node $v_0$ and final node $v_K$; \\
Initial Dirichlet parameters $\alpha^{(0)}$; \\
Per-$t\bar{a}la$ $n$-gram priors $P_{\mathrm{prior}}(s \mid \tau, \cdot)$, $t\bar{a}la$ prior $P(\tau)$, 
  for all $\tau \in \mathcal{T}$, and transition counts $C(r \to q)$; \\
Hyperparameters $(\rho, \beta, W, \lambda , \Delta_{\mathrm{beam}})$
\Ensure 
Optimal stroke sequence $\hat{\mathbf{s}} = (\hat{s}_1, \hat{s}_2, \ldots, \hat{s}_K)$
\State Initialize expanded lattice: 
  $\mathcal{L}_{\mathrm{out}} \leftarrow \emptyset$

\State Initialize priority queue: $Q \leftarrow \{(v_0, \alpha^{(0)})\}$

\While{Q is not empty}
  \State $(v, \alpha) \leftarrow Q.\text{pop}()$

  \State Retrieve incoming stroke $s_{k-1}$:
    \quad If $v = v_0$: $s_{k-1} = \langle s \rangle$; 
    \quad else: from arc label

    \State Compute ${P_{\mathrm{adaptive}}}(\cdot \mid s_{k-1})$ 
      \hfill Eq.~\eqref{eq:P_adaptive} 

\If{Bayesian global prior is used}
    \State Estimate $t\bar{a}la$ posterior:
    $P(\tau \mid s_{k-W:k-1})$
    \hfill Eq.~\eqref{eq:P_tau_h}

    \State Compute global prior:
    $P_{\mathrm{prior}}^{(\mathrm{TI})}(\cdot \mid s_{k-W:k-1})$
    \hfill Eq.~\eqref{eq:P_prior_TI}

\Else
    \State Compute global prior directly using neural model:
    $P_{\mathrm{NN}}(\cdot \mid s_{1:k-1})$
\EndIf

    \State Compute combined probability: 
      $P_{\mathrm{comb}}(\cdot \mid s_{k-W:k-1})$ 
      . \hfill Eq.~\eqref{eq:P_comb} 

\State Compute acoustic confidence $C_k$ from outgoing arcs of $v$
  \hfill Eq.~\eqref{eq:confidence}

    \ForAll{outgoing arc $e=(v \rightarrow v')$ labeled $q \in \mathcal{S}$}

    \State Compute rescored arc weight: 
  $\mathrm{score}(e)$ 
  \hfill Eq.~\eqref{eq:score_e}

        \State Update Dirichlet parameters: 
          $\alpha'$ 
          \hfill Eq.~\eqref{eq:dirichlet_update}

    \State Create expanded state $(v', \alpha')$ with cumulative score: 
      $\mathrm{score}(v,\alpha) + \mathrm{score}(e)$ and add to $\mathcal{L}_{\mathrm{out}}$

    \State Push $(v', \alpha')$ to priority queue $Q$
    \EndFor

  \State Apply beam pruning: 
      Retain states in $Q$ with $\mathrm{score}(v,\alpha) \ge \mathrm{score}_{\max} - \Delta_{\mathrm{beam}}$
      \hfill (Eq.~\ref{eq:beam_p})
\EndWhile

  \State  Retrieve best-first search result:  
  Find the best final expanded state 
  $(v_K, \alpha_K) \leftarrow \arg\max_{(v,\alpha) \in \mathcal{L}_{\mathrm{out}}} \mathrm{score}(v,\alpha)$

\State Reconstruct optimal sequence $\hat{\mathbf{s}}$ by backtracking 
  from $(v_K, \alpha_K)$ to $(v_0, \alpha^{(0)})$

\State \Return $\hat{\mathbf{s}}$

\end{algorithmic}
\end{algorithm}

Algorithm~\ref{algo_1} summarizes the proposed offline lattice-rescoring procedure using best-first search with beam pruning. The expanded states $(v,\alpha)$ preserve path-specific rhythmic context, and the optimal stroke sequence is recovered by backtracking from the best final state. An illustrative example is provided in Supplementary Material (S. VIII, Fig. S3).


\section{Curated Tabla Dataset}\label{curated_dataset}
Annotated datasets for Indian percussion, particularly for tabla performance,
remain extremely limited. Existing datasets are often small, focus on isolated
strokes, or are restricted to a single $t\bar{a}la$ (most commonly
$t\bar{i}nt\bar{a}l$), and rarely capture extended tabla performances. In contrast to
Western music, Hindustani music lacks large-scale annotated corpora
suitable for computational research, especially for sequence-level modelling.
To address this limitation, we curate two complementary datasets:
(A){ a performance-recorded theka-centered tabla dataset with improvisatory variations},
termed the \emph{Tabla Improvisation Dataset (TID)}, and
(B) a \emph{Synthetic Tabla Improvisation Dataset (TID-S)} constructed from
isolated stroke {recordings using expert-defined symbolic stroke sequences}.

\subsection{Tabla Improvisation Dataset (TID)}
The dataset was collected in a small acoustically treated auditorium using a Samsung Galaxy M35 5G smartphone with the native Samsung Voice Recorder application. All recordings were performed by an expert tabla player on a fixed tabla set tuned to
$A\#$ (466.16 Hz).
The performances span tempos from 110 to 240 BPM, covering both $madhya$ and
$drut\ laya$. The recordings primarily consist of canonical $theka$ patterns together with performer-specific variations and short improvisatory passages, rather than full-length traditional tabla solo concert forms.

The dataset comprises approximately 133 minutes of audio distributed
across 23 recordings and covers six commonly used $t\bar{a}la$s:
$Dadr\bar{a}$, $Ekt\bar{a}l$, $Jhapt\bar{a}l$, $Keherv\bar{a}$, $R\bar{u}pak$, and
$T\bar{i}nt\bar{a}l$. The stroke vocabulary contains 30 distinct strokes (listed in Supplementary Material (S. I)), including both basic articulations (elementary single-stroke events)
and compound strokes involving simultaneous gestures on the
dayan (right-hand treble drum) and bayan (left-hand bass drum).
Audio was recorded in stereo at a sampling rate of 44.1 kHz. Stroke
sequence annotations were provided by the performing artist without onset-level
timing information, making the dataset suitable for weakly supervised learning.

\subsection{Synthetic Tabla Improvisation Dataset (TID-S)}
To complement the real-world recordings, we constructed a synthetic tabla dataset from { expert-defined symbolic stroke sequences and isolated stroke recordings. The symbolic sequences consist primarily of canonical $theka$ patterns together with rhythmically valid improvisatory variations and fillers while preserving the underlying $t\bar{a}la$ structure. Given a selected tempo, a Weighted Finite-State Transducer (WFST) -based stroke-sequence generator assigns onset times to the symbolic stroke events, after which waveform instances are selected from a library of isolated stroke recordings and combined to synthesize the final audio. The isolated stroke recordings were obtained from the same performer, using the same tabla set and recording setup as TID}. The synthetic corpus has a total duration of approximately 121 minutes and uses the same 30-stroke vocabulary as TID. 

Together, TID provides acoustic realism, while TID-S provides controlled rhythmic diversity for systematic evaluation across multiple $t\bar{a}la$s.

\section{Experimental Setup}\label{sec_exp_setup}
The experimental setup follows the framework introduced in
Section~\ref{sec_proposed_system}. We report results under two decoding
conditions: (i) acoustic-only decoding, which serves as the baseline, and
(ii) rhythmically rescored decoding using the proposed ADRM. In addition,
we include strongly supervised baseline systems to provide an upper bound
on performance and conduct ablation studies to analyze the contributions
of individual components within the ADRM framework.

\subsection{Datasets}
We evaluate the proposed method using four datasets: two curated datasets
introduced in this work—the TID and the TID-S—together with {two publicly
available benchmarks, the Tabla Solo Dataset (TSD)~\cite{S8}, which contains
tabla solo recordings, and the Hindustani Music Rhythm Dataset (HMR)~\cite{G4},
which contains tabla accompaniment excerpts from full concert performances.
Although the public datasets provide stroke annotations aligned to onset times,
we discard the onset information. We retain only symbolic stroke sequences to
ensure a consistent weakly supervised setting across all datasets.

Each dataset is randomly partitioned into training, validation (for hyperparameter tuning),
and test sets, ensuring that all stroke classes are represented across the splits. Training sets are augmented using pitch and time scaling, attack remixing, spectral filtering, and stroke remixing, as in~\cite {S9}. Each augmentation
method generates one additional augmented version of the original training
set. For HMR, augmentation 
is omitted because the recordings contain vocals and accompaniment.
Only original (non-augmented) recordings are used for
validation and testing. Dataset statistics are summarized in Table~\ref{T_dataset}.

\subsection{Acoustic Model and Training Configuration}
\label{sec:acoustic_model}
We adopt the \emph{k2} finite-state framework together with the Icefall
toolkit\footnote{https://github.com/k2-fsa/icefall} to train a CTC-based
acoustic model and generate decoding lattices. The acoustic model is based on a Time-Delay Neural Network (TDNN), chosen
for its efficient temporal modelling and parameter-efficient factorization.
Specifically, we employ a convolutional, factorized TDNN (C-TDNN-F)
architecture, building on the factorized TDNN formulation based on
low-rank matrix factorization~\cite{Povey2018} and prior convolutional
TDNN-F architectures reported for speech recognition~\cite{Psutka2021}.

The acoustic model operates on log-Mel spectrogram features (46.4~ms
analysis window, 10~ms shift, 128 Mel filter banks) normalized per
frequency. The architecture uses 12 C-TDNN-F layers. Each layer includes
a bottleneck projection, temporal convolution (context $[-1,0,+1]$),
and residual connections. Batch normalization and dropout (rate 0.1)
are applied throughout. Increasing dilation factors expands the
effective receptive field.

Training is performed using the CTC objective defined in
Eq.~(\ref{eq:CTC}) within the \emph{k2} framework. The model is optimized
using Adam with a learning rate of 0.001 for 100 epochs, with gradient
clipping applied to stabilize training.


\subsection{Hyperparameter Settings}
\label{sec:hyperparameters}
The ADRM framework introduces hyperparameters controlling rhythmic memory ($\rho$), acoustic--rhythmic weighting ($\beta,\lambda$), and the beam search procedure. We perform a hyperparameter search over
$\rho \in \{0.05, 0.03, 0.02\}$, corresponding to approximate effective
memories of 20, 33, and 50 stroke updates, respectively. $\beta \in \{0, 0.2, 0.4, 0.6, 0.8, 1.0\}$  and $\lambda \in \{0, 0.2, 0.4, 0.6, 0.8, 1.0\}$. A grid search is performed independently for each dataset over all combinations of $\rho$, $\beta$, and $\lambda$ to identify the optimal hyperparameter configuration for the confidence-modulated ADRM based on validation-set SER. The search yields $\rho = 0.03$, $\beta = 0.6$, and $\lambda = 0.4$.
For the ablation studies, the remaining hyperparameters are fixed, and only the parameter under investigation is varied. The stroke-history context window is fixed at $W=16$ and is used for
$t\bar{a}la$ posterior estimation and global rhythmic prior computation. The window comprises 16 preceding stroke symbols and is not
imposed as a fixed $t\bar{a}la$-cycle length.}
 Beam search parameters are fixed across all experiments with
$k_{\mathrm{beam}} = 150$ and $\Delta_{\mathrm{beam}} = 10$. Smoothing constants
are fixed to $\varepsilon_{\mathrm{dir}} = 1.0$ (Dirichlet initialization,
Eq.~\ref{eq:alpha_init}). These parameters were selected
heuristically.

\subsection{Decoding and Lattice Generation}
\label{sec:decoding}

After training, decoding is performed using beam search to generate full CTC decoding lattices. For acoustic-only evaluation, the final stroke sequence is obtained by Viterbi decoding over the acoustic lattice using only acoustic scores. For rhythmic rescoring, the same CTC lattice is provided to the state-expansion procedure summarized in Algorithm~\ref{algo_1}. Expanded states $(v,\alpha)$ preserve path-specific rhythmic context, and the final stroke sequence is obtained from the highest-scoring path after rhythmic rescoring and beam pruning.

\subsection{Neural Next-Stroke Language Model (NN-LM) Training}
\label{sec:NN_LM_training}

The NN-LM is trained on the same ground-truth stroke sequence partitions used for the Bayesian global rhythmic prior. {Each stroke symbol is mapped to a trainable embedding vector with dimensionality $d = 64$. The embedding parameters are jointly learned with the NN-LM using the next-stroke prediction objective. The resulting embeddings are processed by a two-layer bidirectional GRU with 256 hidden units per direction, followed by a linear projection and softmax normalization over the dataset-specific stroke vocabulary $S$.} Training uses a next-stroke prediction objective optimized with Adam (learning rate 0.001), with early stopping based on validation set loss 
to prevent overfitting. No data augmentation is applied to symbolic 
stroke sequences. After training, the NN-LM serves as the global 
rhythmic prior component within the ADRM framework during lattice 
rescoring and hyperparameter tuning (Section~\ref{sec:hyperparameters}).

\subsection{Baseline Systems}
For a fair comparison, all baseline systems are evaluated in a common symbolic
stroke-sequence representation using the same sequence-level Stroke Error Rate
(SER). The core methodology of each published baseline is retained, with
task-specific adaptations made only where required for the present TST
evaluation. We consider the fully supervised OTM~\cite{S10}, PTM2~\cite{meta},
and Segment-and-Classify (S\&C)~\cite{S12} approaches, together with the
unsupervised stroke-clustering framework
~\cite{1_4_gowriprasad2025unsupervised}. For the unsupervised
baseline, the original processing and clustering procedure is retained, and
the resulting acoustic clusters are mapped to the 30-stroke vocabulary used
in this study for SER evaluation. OTM, PTM2, and S\&C are evaluated on TSD
and HMR, whereas the unsupervised baseline is evaluated on all four datasets.
Implementation details are provided in Supplementary Material (S.~II--S.~III).

\subsection{Temporal Alignment Evaluation}
To assess whether temporal information can be recovered from the proposed weakly supervised framework, we perform a preliminary temporal alignment analysis on the TSD and HMR test sets, which provide onset-level annotations. After ADRM rescoring, the final stroke sequence is aligned to the frame-level posterior probabilities produced by the CTC acoustic model using CTC segmentation \cite{CTC_segment}. Alignment quality is evaluated by comparing the recovered onset times against the ground-truth stroke onsets. Following standard MIR practice, onset F1-scores are reported using tolerance windows of 50ms, 100ms, 200ms, and 500ms.

\subsection{Ablation Studies}
\label{sec:ablation_study}

We conduct additional ablation studies to analyze three aspects of the proposed ADRM framework: (i) the effect of training-data size on global rhythmic priors, (ii) the contribution of global and local rhythmic components through the interpolation parameter $\lambda$, and (iii) adaptive versus fixed acoustic--rhythmic weighting during decoding. Detailed experimental protocols, parameter settings, and results are provided in Supplementary Material (S. IV–S. V).

\begin{table*}[t]
\centering
\begin{minipage}[t]{0.24\textwidth}
\centering
\caption{Dataset Statistics.}
\label{T_dataset}
\setlength{\tabcolsep}{3pt}
\begin{tabular}{@{}lccc@{}}
\toprule
\textbf{Dataset} &
\makecell[c]{\textbf{Train}\\\textbf{(min)}} &
\makecell[c]{\textbf{Val.}\\\textbf{(min)}} &
\makecell[c]{\textbf{Test}\\\textbf{(min)}} \\ \midrule
\textbf{TSD~\cite{S8}} & 10 + 40* & 4 & 4 \\[1mm]
\textbf{HMR~\cite{G4}} & 212 & 60 & 30 \\
\textbf{TID}   & 93 + 372* & 20 & 20 \\
\textbf{TID-S} & 81 + 324* & 20 & 20 \\
\bottomrule
\end{tabular}
\begin{tablenotes}
\item * indicates augmented data.
\end{tablenotes}
\end{minipage}
\hfill
\begin{minipage}[t]{0.74\textwidth}
\centering
\caption{SER (\%) comparison across TST systems.}
\vspace{-2mm}
\label{tab:main_single}
\setlength{\tabcolsep}{2.5pt}

\begin{tabular}{@{}l ccc c c cc@{}}
\toprule
\textbf{Data} &
\multicolumn{3}{c}{\textbf{Strongly Supervised}} &
\textbf{Unsupervised} &
\textbf{Acoustic} &
\textbf{Acoustic + ADRM} &
\textbf{Acoustic + ADRM} \\
\cmidrule(lr){2-4}
&
\textbf{OTM\cite{S10}} &
\textbf{PTM2\cite{meta}} &
\textbf{S\&C\cite{S12}} &
\textbf{Clustering \cite{1_4_gowriprasad2025unsupervised}} &
\textbf{Only} &
\textbf{Global Bayesian Prior} &
\textbf{Global Neural Prior} \\
\midrule
\textbf{TSD}   &
27.7 &
24.5 &
40.2 &
59.3 &
38.6 &
31.3 (↓18.9\%) &
33.8 (↓12.4\%) \\

\textbf{HMR}   &
28.5 &
26.2 &
67.4 &
84.1 &
43.8 &
29.9 (↓31.7\%) &
29.6 (↓32.4\%) \\

\textbf{TID}   &
-- &
-- &
-- &
47.4 &
24.9 &
16.4 (↓34.1\%) &
17.5 (↓29.7\%) \\

\textbf{TID-S} &
-- &
-- &
-- &
42.3 &
17.8 &
11.5 (↓35.4\%) &
11.7 (↓34.3\%) \\
\bottomrule
\end{tabular}

\begin{tablenotes}
\item Parenthesized values denote relative SER reduction (↓) over
acoustic-only decoding.
\item ``--'' denotes not applicable, since the required annotations for
training these baselines are unavailable for these datasets.
\end{tablenotes}
\end{minipage}
\end{table*}
\section{Results and Discussion}
\label{sec_results}

Performance is evaluated using \emph{Stroke Error Rate (SER)}, analogous to Word Error Rate (WER) in ASR. SER measures transcription errors through substitutions ($S$), deletions ($D$), and insertions ($I$), and is computed using the Levenshtein edit distance between the reference and predicted stroke sequences:
\begin{equation}
\mathrm{SER} = \frac{S + D + I}{N},
\end{equation}
where $N$ is the number of strokes in the reference sequence. In the results tables, “Global Bayesian Prior” and “Global Neural Prior” refer to the models described in Sections~\ref{static_prior} and~\ref{sec:NN_LM_training}, respectively. The “Acoustic Only” baseline uses CTC decoding without rhythmic rescoring.

Table~\ref{tab:main_single} summarizes the results obtained using the selected
hyperparameters $\rho=0.03$, $\beta=0.6$, and $\lambda=0.4$
(Section~\ref{sec:hyperparameters}), applied uniformly across all datasets.
The proposed Acoustic Model + ADRM consistently outperforms
acoustic-only decoding, with relative SER reductions of 18.9--35.4\% across
the four datasets. The Bayesian prior performs best on TSD, TID, and TID-S, while the neural prior is slightly better on HMR.

The proposed framework also outperforms the fully supervised
S\&C baseline on both TSD and HMR, despite S\&C using explicit onset-based
segmentation. The unsupervised
stroke-clustering baseline gives the highest SER on all four datasets.
The advantage over S\&C arises from the sequence-based nature of CTC,
where stroke predictions can benefit from the preceding stroke sequence and
surrounding temporal context, rather than being determined only after
separately detected onsets. The lower SER than the unsupervised clustering
baseline reflects the benefit of the additional sequence-level target
information available during weakly supervised training.

Performance varies across datasets. TSD is constrained by the limited amount
of original training data, despite relatively lesser interference from vocals
or accompanying instruments, whereas HMR contains 151 concert excerpts from
45 artists with greater interference from vocals and accompanying instruments,
spanning four $t\bar{a}las$. The observed $m\bar{a}tr\bar{a}s$-per-minute values also
vary substantially within the same $t\bar{a}la$, ranging from 21.85--372.67 for
$T\bar{i}nt\bar{a}l$ and 10.35--301.51 for $Ekt\bar{a}l$. These variations reflect differences in performance tempo and temporal realization rather than changes in the underlying cyclic metrical organization of the corresponding $t\bar{a}la$. Despite these challenges, the
proposed framework (Acoustic Model + ADRM) achieves a 31.7\% relative SER
reduction on HMR, indicating that rhythmic modeling remains beneficial under
these conditions. The larger TID and TID-S datasets show further relative
SER reductions of 34.1--35.4\%. Representative Mel-spectrograms illustrating
the different spectro-temporal characteristics of these datasets are shown
in Fig.~S2 of the Supplementary Material.

The ablation results (Supplementary Material (S. V)) further support the proposed ADRM design. The Bayesian global prior is more effective when limited symbolic rhythmic training data are available, whereas the neural prior becomes increasingly competitive as training data increase. Experiments with different interpolation weights show that combining global and local rhythmic information yields lower SER than using either component alone, with the best performance consistently obtained at $\lambda=0.4$. In addition, confidence-modulated rhythmic weighting provides small but consistent improvements over fixed-weight decoding across datasets. 

{
\begin{table}[h]
\centering
\caption{Temporal alignment performance (\% f1-score).}
\label{tab:temporal_alignment}
\begin{tabular}{lcccc}
\toprule
\textbf{Dataset} & \textbf{f1@50 ms} & \textbf{f1@100 ms} & \textbf{f1@200 ms} & \textbf{f1@500 ms} \\
\midrule
TSD & 8.9 & 11.4 & 20.2 & 36.4 \\
HMR & 6.2 & 9.8 & 17.1 & 32.3 \\
\bottomrule
\end{tabular}
\end{table}
Table~\ref{tab:temporal_alignment} reports temporal alignment performance obtained using CTC segmentation. Although onset f1-scores are relatively low under strict tolerance windows, performance consistently improves as the tolerance window increases for both datasets. This suggests that the frame-level representations learned by the weakly supervised CTC model retain recoverable temporal information despite the absence of onset-level supervision during training. However, accurate onset localization remains challenging, indicating that explicit temporal modelling remains an important direction for future work}.

\subsection{Analysis of Global and Adaptive Rhythmic Modelling}
\label{sec:adrm_analysis}
\begin{figure*}[t]
    \centering
    \includegraphics[width=\textwidth]{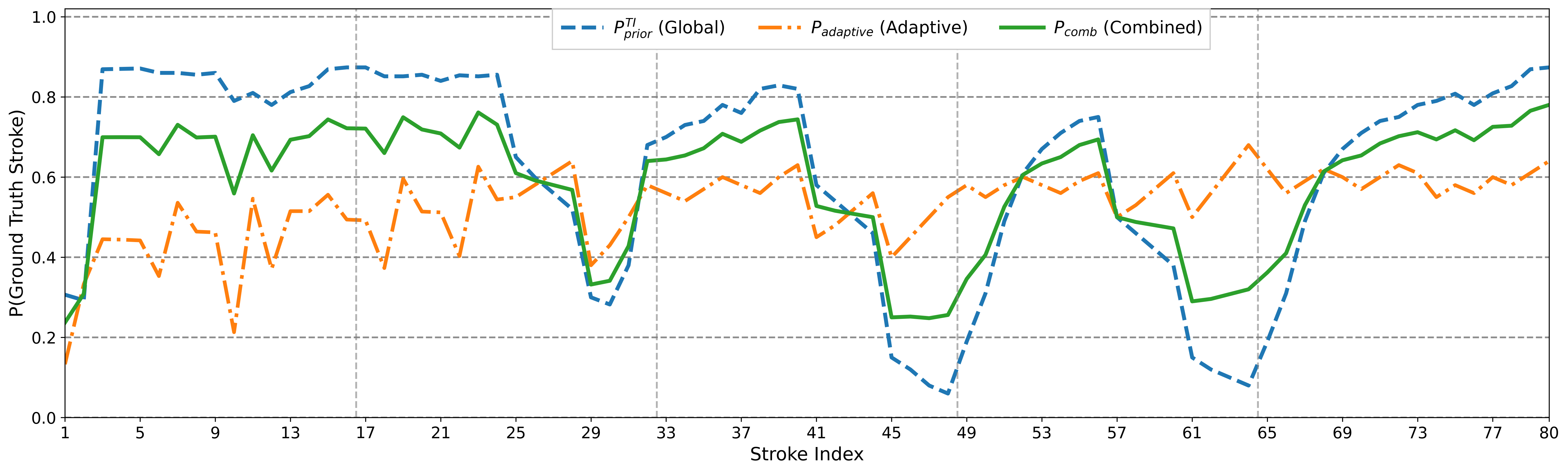}
\caption{Evolution of the global rhythmic prior ($P^{(TI)}_{\mathrm{prior}}$),
local adaptive rhythm model ($P_{\mathrm{adaptive}}$), and their interpolation
for a representative 80-stroke sequence spanning five consecutive
$T\bar{i}nt\bar{a}l$ cycles. Dashed lines indicate illustrative cycle
boundaries used only to organize the analysis; cycle boundaries are not
provided as inputs to ADRM.}

    \label{fig:ADRM_analysis}
\end{figure*}
To further examine the behaviour of ADRM, Fig.~\ref{fig:ADRM_analysis} shows the evolution of $P_{\mathrm{prior}}^{\mathrm{TI}}$, $P_{\mathrm{adaptive}}$, and $P_{\mathrm{comb}}$ for a representative 80-stroke sequence spanning five consecutive $tint\bar{a}l$ cycles. 
The complete stroke sequence together with a detailed cycle-wise description is provided in the Supplementary Material (S. VI). 
For this analysis, the stroke sequence is supplied directly to ADRM and the probabilities are computed sequentially at each stroke position. Consequently, the observed probability trajectories reflect the behaviour of the rhythmic models themselves rather than acoustic-model or decoding effects.

The probability trajectories reveal distinct roles for $P_{\mathrm{prior}}^{\mathrm{TI}}$ and $P_{\mathrm{adaptive}}$. During the canonical portions of the sequence, both probabilities remain relatively high. As the sequence progresses from the \emph{Dhage Traka Tin Ta} region in Cycle~2 to the \emph{Traka Dhin Dhin Dha} and subsequently the repeated \emph{Tita Kata Gadi Gana} phrases in Cycles~3 and~4, $P_{\mathrm{prior}}^{\mathrm{TI}}$ exhibits progressively larger reductions, whereas $P_{\mathrm{adaptive}}$ varies within a comparatively narrower range. The repeated \emph{Tita Kata Gadi Gana} phrase further highlights this difference: while the corresponding $P_{\mathrm{prior}}^{\mathrm{TI}}$ values remain low across both occurrences, the adaptive probabilities increase during the second occurrence, reflecting the influence of the recently observed local pattern. Because these updates are driven by recently observed transitions, future predictions can become increasingly influenced by the accumulated local history.

In contrast, $P_{\mathrm{prior}}^{\mathrm{TI}}$ remains largely unchanged between the two occurrences of the \emph{Tita Kata Gadi Gana} phrase because it is determined by the learned global rhythmic model rather than the recent stroke history. Consequently, the global prior provides a stable reference derived from $t\bar{a}la$-conditioned stroke-sequence statistics, but it does not adapt to recurring local patterns within the ongoing sequence. The adaptive model addresses this limitation by updating its predictions as new stroke transitions are observed. Thus, the two components serve complementary roles: $P_{\mathrm{prior}}^{\mathrm{TI}}$ provides stability, whereas $P_{\mathrm{adaptive}}$ provides adaptability.

These differences are reflected in the evolution of $P_{\mathrm{comb}}$. In the latter part of Cycles~3 and~4, where the divergence between $P_{\mathrm{prior}}^{\mathrm{TI}}$ and $P_{\mathrm{adaptive}}$ is most pronounced, the combined probability follows neither component individually.
Instead, it incorporates information from both the stable $t\bar{a}la$-specific prior and the evolving local transition history, allowing ADRM to remain responsive to recurring local patterns while retaining a stable rhythmic reference.

The trends observed in Fig.~\ref{fig:ADRM_analysis} are consistent with the $\lambda$-ablation results presented in Table~SI (Supplementary Material). The figure shows that the global rhythmic prior and the local adaptive model contribute complementary information during prediction. Accordingly, the lowest SER is obtained at $\lambda=0.4$, where both components are jointly utilized.

\subsection{Error Analysis}
To better understand the behaviour of the proposed framework, we analyzed the transcription errors by decomposing the Stroke Error Rate (SER) into substitution (S), deletion (D), and insertion (I) components and by examining the dominant stroke-level error patterns before and after ADRM rescoring. The corresponding dataset-wise statistics are provided in Supplementary Material (S. VII).

Across all datasets, substitution errors constitute the largest component of the SER. In TID, the most frequent substitutions are \textit{Dhin}$\rightarrow$\textit{Dha}, \textit{Tin}$\rightarrow$\textit{Na}, \textit{Ti}$\rightarrow$\textit{Ta}, and \textit{Dha}$\rightarrow$\textit{Dhin}. In TSD, the dominant substitutions are \textit{Dhin}$\rightarrow$\textit{Dhi}, \textit{Dha}$\rightarrow$\textit{Da}, and \textit{Tin}$\rightarrow$\textit{Ta}, whereas the HMR dataset exhibits similar substitution patterns together with additional confusions that arise in the presence of singing voice and harmonic accompaniment. These observations are consistent with earlier studies on tabla acoustics. These observations are consistent with earlier studies showing that acoustically related tabla strokes tend to form common timbral groups~\cite{S12,1_4_gowriprasad2025unsupervised}. Several of the dominant substitution pairs observed in our experiments belong to these acoustically related groups. 

Insertion and deletion errors show different characteristics from substitution errors. The most frequently inserted strokes are generally among the dominant labels in the corresponding datasets, whereas deletion errors are concentrated on comparatively less frequent stroke classes. This indicates that frequently occurring strokes are more likely to be predicted, whereas strokes with fewer training examples are more likely to be missed. 

The error analysis also provides insight into why ADRM improves transcription accuracy. The largest gains are consistently obtained for insertion errors, whose contribution decreases from 7.7\% to 2.8\% for TID, from 5.7\% to 3.0\% for TSD, and from 9.4\% to 4.0\% for HMR. Since inserted strokes introduce additional transitions that disrupt the local rhythmic structure, they are often assigned lower probabilities during rhythm-aware lattice rescoring. As illustrated in Fig.\ref{fig:ADRM_analysis}, the adaptive model continuously updates transition probabilities based on the decoded stroke history, allowing inconsistent continuations to be suppressed.

Substitution errors show smaller but consistent improvements because the lattice often contains multiple acoustically plausible stroke sequences. ADRM can reorder these competing hypotheses using rhythmic evidence, but acoustically similar strokes may still receive comparable scores. Deletion errors show the smallest improvements because missing strokes often do not survive in the decoding lattice and therefore cannot be recovered during rescoring.

Overall, the error analysis suggests that ADRM effectively exploits rhythmic context to reduce insertion and substitution errors during rescoring. However, errors caused by strong acoustic confusions or missing stroke hypotheses remain challenging to correct.

\section{Conclusion}
\label{sec_conclusion}
 This work presents ADRM, a rhythm-aware decoding framework for weakly supervised Tabla Stroke Transcription that operates without onset-level annotations. By combining CTC-based acoustic modelling with global rhythmic priors and local adaptive rhythmic refinement, ADRM consistently reduces Stroke Error Rate across diverse datasets, demonstrating the benefit of incorporating $t\bar{a}la$-conditioned symbolic rhythmic regularities and local adaptive transition information during weakly supervised TST.

To support this task, we also introduce the Tabla Improvisation Dataset (TID) and its synthetic counterpart (TID-S), providing new benchmarks for weakly supervised tabla transcription. { Ablation studies further highlight the complementary roles of global and local rhythmic components, the data efficiency trade-offs between analytical and learned priors, and the small but consistent SER reductions obtained through confidence-modulated acoustic--rhythmic weighting. A preliminary temporal alignment analysis using CTC-based segmentation further demonstrates the feasibility of recovering approximate onset timings despite training with only sequence-level annotations.}

Future work will focus on improving temporal alignment accuracy, extending ADRM to online decoding with incremental rhythmic adaptation, and adapting the framework to other percussion traditions such as Carnatic mridangam.

\section*{Acknowledgments}

The authors would like to thank Ashwin Sundram (Tabla Visharad Pratham) for performing 
the tabla, curating the real-world tabla dataset, and providing symbolic stroke 
sequence annotations (without temporal alignment) that form the foundation of this work.

\bibliographystyle{IEEEtran}
\bibliography{main}

@inproceedings{CTC,
  title={Connectionist temporal classification: labelling unsegmented sequence data with recurrent neural networks},
  author={Graves, Alex and Fern{\'a}ndez, Santiago and Gomez, Faustino and Schmidhuber, J{\"u}rgen},
  booktitle={ICML},
  pages={369--376},
  year={2006}
}

@article{meta,
  title={Meta-Learning-Based Percussion Transcription and $t\bar{a}la$ Identification From Low-Resource Audio},
  author={Kodag, Rahul Bapusaheb and Arora, Vipul},
  journal={IEEE Transactions on Audio, Speech and Language Processing},
  year={2025},
  publisher={IEEE}
}

@article{D4,
  title={A review of automatic drum transcription},
  author={Wu, Chih-Wei and Dittmar, Christian and Southall, Carl and Vogl, Richard and Widmer, Gerhard and Hockman, Jason and M{\"u}ller, Meinard and Lerch, Alexander},
  journal={IEEE/ACM Transactions on Audio, Speech, and Language Processing},
  volume={26},
  number={9},
  pages={1457--1483},
  year={2018},
  publisher={IEEE}
}

@inproceedings{G4,
  title={A generalized Bayesian model for tracking long metrical cycles in acoustic music signals},
  author={Srinivasamurthy, Ajay and Holzapfel, Andre and Cemgil, Ali Taylan and Serra, Xavier},
  booktitle={ICASSP},
  pages={76--80},
  year={2016},
  organization={IEEE}
}

@article{G5,
  title={A Data-driven bayesian approach to automatic rhythm analysis of indian art music},
  author={Srinivasamurthy, Ajay and others},
  journal={PhD thesis, Universitat Pompeu Fabra},
  year={2016},
  publisher={Universitat Pompeu Fabra}
}

@inproceedings{S8,
  title={Discovery of syllabic percussion patterns in tabla solo recordings},
  author={Gupta, Swapnil and Srinivasamurthy, Ajay and Kumar, Manoj and Murthy, Hema A and Serra, Xavier},
  booktitle={ISMIR},
  year={2015},
}

@inproceedings{S9,
  title={Four-way classification of tabla strokes with models adapted from Automatic Drum Transcription},
  author={MA, Rohit and Bhattacharjee, Amitrajit and Rao, Preeti},
  booktitle={ISMIR},
  year={2021},
}

@article{S10,
  title={Four-way classification of tabla strokes with transfer learning using western drums},
  author={Ananthanarayana, Rohit M and Bhattacharjee, Amitrajit and Rao, Preeti},
  journal={Transactions of the International Society for Music Information Retrieval},
  volume={6},
  number={1},
  year={2023}
}

@article{S12,
  title={Automatic stroke classification of tabla accompaniment in Hindustani vocal concert audio},
  author={Rao, Preeti and others},
  journal={arXiv preprint arXiv:2104.09064},
  year={2021}
}

@inproceedings{R6,
  title={Unaligned supervision for automatic music transcription in the wild},
  author={Maman, Ben and Bermano, Amit H},
  booktitle={ICML},
  pages={14918--14934},
  year={2022},
  organization={PMLR}
}

@inproceedings{CTC_ADT,
  title={CTC2: End-to-End Drum Transcription Based on Connectionist Temporal Classification With Constant Tempo Constraint},
  author={Kamakura, Daichi and Nakamura, Eita and Yoshii, Kazuyoshi},
  booktitle={Asia Pacific Signal and Information Processing Association Annual Summit and Conference (APSIPA ASC)},
  pages={158--164},
  year={2023},
  organization={IEEE}
}

@article{ASR_review_2025,
  title={Automatic Speech Recognition: A survey of deep learning techniques and approaches},
  author={Ahlawat, Harsh and Aggarwal, Naveen and Gupta, Deepti},
  journal={International Journal of Cognitive Computing in Engineering},
  year={2025},
  publisher={Elsevier}
}

@article{ASR_error_correction_TR_2025,
  title={Asr error correction using large language models},
  author={Ma, Rao and Qian, Mengjie and Gales, Mark and Knill, Kate},
  journal={IEEE Transactions on Audio, Speech and Language Processing},
  year={2025},
  publisher={IEEE}
}

@inproceedings{MLM_2021,
  title={Statistical correction of transcribed melody notes based on probabilistic integration of a music language model and a transcription error model},
  author={Hiramatsu, Yuki and Shibata, Go and Nishikimi, Ryo and Nakamura, Eita and Yoshii, Kazuyoshi},
  booktitle={ICASSP},
  pages={256--260},
  year={2021},
  organization={IEEE}
}

@article{MLM_2024,
  title={Towards efficient and real-time piano transcription using neural autoregressive models},
  author={Kwon, Taegyun and Jeong, Dasaem and Nam, Juhan},
  journal={IEEE/ACM Transactions on Audio, Speech, and Language Processing},
  year={2024},
  publisher={IEEE}
}

@inproceedings{MLM_2025,
  title={A Hybrid Model for Automatic Music Transcription of Hulusi Based on Multi-Feature Fusion},
  author={Li, Zhi and Zhou, Yi and Wu, Hui},
  booktitle={Proceedings of the 4th International Conference on Artificial Intelligence and Intelligent Information Processing},
  pages={30--34},
  year={2025}
}

@article{1_4_gowriprasad2025unsupervised,
  title={Unsupervised stroke clustering for developing common stroke set for percussion in Indian art music},
  author={Gowriprasad, R and Kuriakose, Jom and Aravind, R and Murthy, Hema A and others},
  journal={The Journal of the Acoustical Society of America},
  volume={158},
  number={5},
  pages={3711--3723},
  year={2025},
  publisher={AIP Publishing}
}

@inproceedings{1_gowriprasad2020onset,
  title={Onset detection of tabla strokes using lp analysis},
  author={Gowriprasad, R and Murty, K Sri Rama},
  booktitle={2020 International Conference on Signal Processing and Communications (SPCOM)},
  pages={1--5},
  year={2020},
  organization={IEEE}
}

@article{4_gowriprasad2022structural,
  title={Structural segmentation and labelling of tabla solo performances},
  author={Gowriprasad, R and Aravind, R and Murthy, Hema A},
  journal={Journal of New Music Research},
  volume={51},
  number={4-5},
  pages={300--320},
  year={2022},
  publisher={Taylor \& Francis}
}

@inproceedings{4_gowriprasad2021tabla,
  title={Tabla Gharana Recognition from Audio music recordings of Tabla Solo performances.},
  author={Gowriprasad, R and Venkatesh, V and Murthy, Hema A and Aravind, R and Murty, K Sri Rama},
  booktitle={ISMIR},
  pages={547--554},
  year={2021}
}

@inproceedings{mridang,
  title={Akshara transcription of mrudangam strokes in carnatic music},
  author={Kuriakose, Jom and Kumar, J Chaitanya and Sarala, Padi and Murthy, Hema A and Sivaraman, Umayalpuram K},
  booktitle={2015 Twenty First National Conference on Communications (NCC)},
  pages={1--6},
  year={2015},
  organization={IEEE}
}

@article{Povey2018,
  title={Semi-orthogonal low-rank matrix factorization for deep neural networks},
  author={Yarmohammadi, Mahsa},
  journal={Interspeech 2018},
  year={2018}
}

@inproceedings{Psutka2021,
  title={CNN-TDNN-based architecture for speech recognition using grapheme models in bilingual czech-slovak task},
  author={Psutka, Josef V and {\v{S}}vec, Jan and Pra{\v{z}}{\'a}k, Ale{\v{s}}},
  booktitle={International Conference on Text, Speech, and Dialogue},
  pages={523--533},
  year={2021},
  organization={Springer}
}

@inproceedings{sankaran2015,
  title={Automatic segmentation of composition in carnatic music using time-frequency cfcc templates},
  author={Sankaran, Sridharan and Krishnaraj Sekhar, PV and Hema, A Murthy},
  booktitle={Proceedings of 11th international symposium on computer music multidisciplinary research},
  year={2015}
}

@inproceedings{CTC_segment,
  title={Ctc-segmentation of large corpora for german end-to-end speech recognition},
  author={K{\"u}rzinger, Ludwig and Winkelbauer, Dominik and Li, Lujun and Watzel, Tobias and Rigoll, Gerhard},
  booktitle={International Conference on Speech and Computer},
  pages={267--278},
  year={2020},
  organization={Springer}
}

\end{document}